\documentclass[runningheads]{llncs}

\usepackage{amsmath} \usepackage{amscd} \usepackage{amssymb} \usepackage{amsfonts} \usepackage{array}
\usepackage{graphics}
\usepackage[all]{xy}
\usepackage{color}

\newtheorem{assumption}{Assumption}

\newcommand{\msf}[1]{{\mathsf{#1}}}
\newcommand{\mbf}[1]{{\mathbf{#1}}}
\newcommand{\mi}[1]{{\mathit{#1}}}

\begin{document}

\title{How to prevent type-flaw attacks on security protocols under algebraic properties}
\titlerunning{Preventing type-flaw attacks under algebraic properties}

\author{Sreekanth Malladi\inst{1}\thanks{Supported by a doctoral SEED grant from the graduate school at DSU in 2007.}  \and Pascal Lafourcade\inst{2}\thanks{Supported by ANR SeSur SCALP, SFINCS, AVOTE projects.}}
\authorrunning{Malladi, Lafourcade}

\institute{
		Dakota State University \\	
		Madison, SD 57042, USA \\
		\email{Sreekanth.Malladi@dsu.edu}
\and
		Universit\'e Grenoble 1, CNRS,\textsc{Verimag}, \\ 
	  2 avenue de Vignate 38000 Gi\`eres France \\ 						  	\email{pascal.lafourcade@imag.fr}
}

\maketitle

\begin{center}  \today  \end{center}

\begin{abstract}

Type-flaw attacks upon security protocols wherein agents are led to misinterpret message types have been reported frequently in the literature. Preventing them is crucial for protocol security and verification. Heather et al. proved that tagging every message field with it's type prevents all type-flaw attacks under a free message algebra and perfect encryption system. 

~~~~~~~~~~~~~In this paper, we prove that type-flaw attacks can be prevented with the same technique even under the {\sf ACUN} algebraic properties of \texttt{XOR} which is commonly used in ``real-world" protocols such as \texttt{SSL 3.0}. Our proof  method is general and can be easily extended to other  monoidal operators that possess properties such as Inverse and Idempotence as well. We also discuss how tagging could be used to prevent type-flaw attacks under other properties such as associativity of pairing, commutative encryption, prefix property and homomorphic encryption. 

\end{abstract}

\begin{keywordname}
	Cryptographic protocols, Type-flaw attacks, Tagging, Algebraic properties, Equational theories, Constraint solving, Decidability.
\end{keywordname}

\section{Introduction}

A {\em type-flaw attack} on a protocol is an attack where a message variable of one type is essentially substituted with a message of a different type, to cause a violation of a security property.

In their pioneer work, Heather et al. proved that pairing constants called ``tags" with each message prevents type-flaw attacks \cite{HLS03}.

Does preventing type-flaw attacks have advantages?

\begin{itemize}

	\item As Heather et al. pointed out, besides the obvious 							advantage to security in preventing these commonly and frequently reported attacks, preventing 								them also allows many unbounded 										verification approaches (e.g. \cite{THG98,Cohen00,HS00}) 				 to be meaningful, since they 						assume the 							absence of type-flaw attacks;
	\item Further, Ramanujam-Suresh  found that the absence of any 				 type-flaw attacks allows us to restrict analysis to 							well-typed runs only \cite{RS-FST03}, which is a 								decidable problem; i.e., security can be decided with 					analyzing just a single session.

\end{itemize}

Thus, prevention of type-flaw attacks is a crucial and significant result toward protocol analysis and verification.

However, Heather et al.'s work only considered a basic protocol model with a free message algebra and perfect encryption. Operators such as Exclusive-OR and ciphers such as CBC possess algebraic properties that violate these assumptions. Recent focus in research projects world-wide has been to extend protocol analysis with algebraic properties to accommodate ``real-world" protocols (e.g. \cite{KustersT08,EscobarMM07}). Naturally, a corresponding study into type-flaw attacks would be both crucial and interesting.

With this motivation, we examined several algebraic properties described in the survey by Cortier et al. \cite{CDL06} such as:

\begin{itemize}
	\item Associative pairing, Commutative encryption,  and 					Monoidal theories that violate the free algebra 						assumption;
	\item the Prefix property, Homomorphic encryption, and 						Low-exponent RSA weakness that violate the  								perfect encryption assumption.
\end{itemize}

We report our observations in this paper. As our main contribution, we prove that type-tagging prevents all type-flaw attacks under \texttt{XOR} that possesses {\sf ACUN} properties (Associativity, Commutativity, existence of Unity and Nilpotence). The proof approach is quite general and can be easily extended to other monoidal theories such as Inverse and Idempotence as well. We also advocate some prudent tagging practices to prevent type-flaw attacks under the other algebraic properties mentioned above.

\paragraph*{Organization.} In Section~\ref{s.type-flaw-ex}, we show how type-tagging can prevent type-flaw attacks under \texttt{XOR} using an example. In  Section~\ref{s.type-flaw-proof}, we give a formal treatment of type-flaw attacks in a symbolic model and provide a simpler  proof compared to \cite{HLS03} that tagging prevents type-flaw attacks under \texttt{XOR}. In Section~\ref{s.others}, we examine how the result withstands each algebraic property and suggest remedies in the form of prudent engineering principles. We sum up with a Conclusion.

\section{Tagging prevents type-flaw attacks under \texttt{XOR} - Example}\label{s.type-flaw-ex}

Consider the adapted Needham-Schroeder-Lowe protocol ~~($\msf{NSL}_{\oplus}$) ~~by Chevalier et al. \cite{CKRT03}:\\

\begin{center}
\begin{tabular}{|l|}
\hline

\parbox[3,5]{3in}{ 

\vskip 0.2in

\begin{center}
\begin{tabular}{ll}

{\bf Msg~~1.}~~$A ~~\to ~~B~~:~~[1,~N_A,~A ]^{\to}_{pk(B)}$ \\

{\bf Msg~~2.}~~$B ~~\to ~~A~~:~~[2,~N_A \oplus B,~N_B ]^{\to}_{pk(A)}$\\

{\bf Msg~~3.}~~$A ~~\to ~~B~~:~~[3,~N_B]^{\to}_{pk(B)}$\\

\end{tabular}
\end{center}

} \\[0.6in]

\hline

\end{tabular}
\end{center}

\vskip 0.1in
($A$ ~~and ~~$B$ ~~are agent variables; ~~$N_A$, ~~$N_B$ ~~are nonce variables; ~~$[X]^{\to}_Y$ ~~represents ~~$X$  ~~encrypted with ~~$Y$ ~~using an asymmetric encryption algorithm.).

A type-flaw attack is possible on this protocol even in the presence of component numbering (recently presented in \cite{MH08}): \\

\begin{center}
\begin{tabular}{|l|}
\hline

\parbox[3,5]{4.5in}{ 

\vskip 0.1in

\begin{center}
\begin{tabular}{ll}

$\mbf{\textcolor{blue}{Msg}} ~~\alpha$$\mbf{.1.}$ ~~$a ~~\to ~~i$ ~~: ~~$[1,~n_a,~a ]_{pk(i)}$ \\

~~~~\textcolor{red}{Msg} ~~$\beta$.1. ~~$i(a) ~~\to  ~~b$ ~~: ~~ $[1,~n_a \oplus b \oplus i,~a ]_{pk(b)}$\\

~~~~\textcolor{red}{Msg} ~~$\beta$.2. ~~$b ~~\to  ~~i(a)$ ~~: ~~ $[2,~n_a \oplus b \oplus i \oplus b,~n_b ]_{pk(a)}$\\

$\mbf{\textcolor{blue}{Msg}} ~~\alpha$$\mbf{.2.}$ ~~$i   ~~\to ~~a$ ~~: ~~$[2,~n_a \oplus i,~n_b]_{pk(a)}$ ~~~~(replaying ~~\textcolor{red}{Msg} ~~$\beta$.2)\\

$\mbf{\textcolor{blue}{Msg}} ~~\alpha$$\mbf{.3.}$ ~~$a    ~~\to ~~i ~~: ~~[3,~n_b ]_{pk(i)}$\\

~~~~\textcolor{red}{Msg} ~~$\beta$.3. ~~$i(a) ~~\to  ~~b$ ~~: ~~$[3,~n_b]_{pk(b)}$\\

\end{tabular}
\end{center} 

} \\[0.6in]

\hline

\end{tabular}
\end{center}

\vskip 0.1in

Notice the type-flaw in the first message ($n_a\oplus b \oplus i$ ~~substituted for the claimed ~~$N_A$) that induces a type-flaw in the second message as well. This is strictly a type-flaw attack since without the type-flaw and consequently without exploiting the algebraic properties, the same attack is not possible.

Component numbering cannot also prevent type-flaw attacks under the ~~{\sf Inverse} ~~property that allows cancellation much like ~~{\sf Nilpotence}. Consider operators ~~$\{+,-\}$, ~~where ~~$+$ ~~is binary addition, ~~$-$ ~~a unary operator, and ~~$0$ ~~a constant. Then, if we change the ~~$\oplus$ ~~operator in the  ~~$\msf{NSL_{\oplus}}$ ~~protocol ~~to ~~$+$, ~~variable ~~$N_A$ ~~could be substituted with ~~$n_a + i - b$ ~~to form the same attack as with ~~$\oplus$.

The above attack can be avoided if type-tagging were to be adopted for the elements of the ~~\texttt{XOR} ~~operator: \\

\begin{center}
\begin{tabular}{|l|}
\hline

\parbox[3,5]{4in}{ 

\vskip 0.1in

\begin{center}
\begin{tabular}{ll}

{\bf Msg ~~1.} ~~$A ~~\to ~~B ~~: ~~[1, ~N_A, ~A]^{\to}_{\mathit{pk}(B)}$ \\

{\bf Msg ~~2.} ~~$B ~~\to ~~A ~~: ~~[ 2, ~[\msf{nonce},N_A] \oplus [\msf{agent}, B], ~N_B ]^{\to}_{\mathit{pk}(A)}$\\

{\bf Msg ~~3.} ~~$A ~~\to ~~B ~~: ~~[3, ~N_B]^{\to}_{pk(B)}$\\

\end{tabular}
\end{center}

} \\[0.4in]

\hline

\end{tabular}
\end{center}

\vskip 0.1in

{\bf \textcolor{red}{Msg}} ~~$\beta${\bf .2} ~~is then not replayable as ~~{\bf \textcolor{blue}{Msg}} ~~$\alpha${\bf .2} ~~even when ~~$i(a)$ ~~sends 
~~{\bf \textcolor{red}{Msg}} ~~$\beta${\bf .1} ~~as ~~$i(a) ~~\to ~~b ~~: ~~[ 1, ~[\msf{nonce},n_a] \oplus [\msf{agent},b] \oplus [\msf{agent},i], ~a ]^{\to}_{\mathit{pk}(b)}$, since ~~{\bf \textcolor{red}{Msg}} ~~$\beta${\bf .2} ~~then becomes ~~$b ~~\to ~~i(a) ~~: ~~[ 2, ~[\msf{nonce}, ~[\msf{nonce}, ~n_a] \oplus [\msf{agent}, ~b] \oplus [\msf{agent}, ~i]] \oplus [\msf{agent}, ~b], ~n_b]^{\to}_{\mathit{pk}(a)}$. 

This is not replayable as the required~~{\bf \textcolor{blue}{Msg}} ~~$\alpha${\bf .2}:  ~~$i  ~~\to  ~~a ~~: ~~[2, ~[\msf{nonce}, ~n_a] \oplus [\msf{agent}, ~i], ~n_b]^{\to}_{\mathit{pk}(a)}$ because, inside ~~{\bf \textcolor{red}{Msg}} ~~$\beta${\bf .2}, ~~one occurence of ~~$[\msf{agent}, ~b]$ ~~is in ~~$[\msf{nonce}, ~[\msf{nonce}, ~n_a] \oplus [\msf{agent}, ~b] \oplus [\msf{agent}, ~i]]$ ~~and the other is outside. Hence, they cannot be canceled.

 A similar reasoning applies to ~~{\sf Inverse} ~~property for ~~$\oplus$ ~~instead of ~~{\sf Nilpotence}. We leave this for the reader to verify.

In the next subsection, we will prove these claims formally.

\section{Type-tagging prevents type-flaw attacks: Proof}\label{s.type-flaw-proof}

In this section, we present a formal proof extending an approach presented in \cite{Mall04} that non-unifiability of encryptions (which can be ensured by tagging with component numbers) prevents type-flaw attacks with free operators and a more detailed type-tagging will prevent them under the monoidal \texttt{XOR} operator. 
Our proof is much simpler than \cite{HLS03}, and more importantly, allows us to easily study extrapolating the result to operators with algebraic properties. Furthermore, being a symbolic protocol model, the framework is quite flexible to include the much needed equational unification  for additional equational theories. 

\subsection{Term Alegbra}\label{ss.term-algebra}

We start off with a term algebra with mostly free operators except for the \texttt{XOR} operator.

\begin{definition}~~{\em \bf [Terms]}\label{d.term-algebra}~~~A \emph{\bf term} is one of the following:

	 \emph{\bf Variable}~~(can be an Agent, Nonce etc., that are all subsets of~~$\mi{Var}$);~~
	 \emph{\bf Constant}~~(numbers~~1,2,~$\ldots$;~~name of the attacker~~$\epsilon$~~etc.);~~
	 \emph{\bf Atom}~~(split into sets agents, nonces etc.);
	 \emph{\bf Concatenation}~~denoted~~$[t_1,~\ldots,~t_n]$~~if~~$t_1,~\ldots,~t_n$~~ are terms;~~
	 \emph{\bf Public-Key}~~denoted~~$\mi{pk}(A)$~~with~~$A$~~of type Agent;~~
	 \emph{\bf Shared-Key}~~denoted~~$\mi{sh}(A,B)$~~with~~$A$~~ and~~$B$~~of type Agent;~~
	 \emph{\bf Asymmetric encryption}~~denoted~~$[t]^{\to}_k$~~ where~~$t$~~and~~$k$~~are terms;~~
	 \emph{\bf Symmetric encryption}~~denoted~~ $[t]^{\leftrightarrow}_k$~~where~~$t$~~and~~$k$~~are terms;
	 ~~\emph{\bf Hash}~~denoted~~$h(t)$~~where~~$t$~~is a term;~~
	 \emph{\bf Signature}~~denoted~~$\mi{Sig}_k(t)$~~where~~$t$~~ is a term to be validated using the key~~$k$;
		~~\emph{\bf XOR} ~~denoted ~~$t_1~\oplus~\ldots~\oplus ~t_n$ ~~where~~$t_1,~\ldots,~t_n$ ~~are terms.
	 
\end{definition}

We will drop the superscript~~$\to$~~ and~~$\leftrightarrow$~~ if the mode of encryption is irrelevant. 

We will call terms with no atoms (but only constants and variables) as ~~{\em parametric terms}. We will call a parametric term in which the variables were substituted with variables and/or atoms as a ~~\emph{semi-term}. 

We will assume that the reader is familiar with the standard definitions of syntactic unification, and the most general unifier (mgu). We will write ~~$t ~~\approx  ~~t'$ ~~if ~~$t$ ~~and ~~$t'$ ~~are unifiable.

As usual,~~{\em subterms}~~are defined to capture parts of messages:

\begin{definition}~~{\em \bf [Subterm]}\label{d.subterm}

Term~~$t$~~is a~~\emph{\bf subterm}~~of~~$t'$~~(denoted~~$t~~ \sqsubset~~ t')$~~if 

\begin{itemize}
	\item $t ~~=~~ t'$, ~~or
	\item $t' ~~= ~~[t_1,\ldots,t_n]$ ~~with ~~$t ~~\sqsubset ~~t^{''}$, where $t^{''}  ~~\in ~~\{ t_1, ~\ldots, ~t_n \}$
 ~~or 
	\item $t' ~~= ~~ [t^{''}]_{k'}$ ~~with ~~$t ~~\sqsubset ~~t^{''}$, ~~or
	\item $t' ~~= ~~h(t^{''})$ ~~with ~~$t ~~\sqsubset ~~t^{''}$, ~~or
	\item $t' ~~= ~~\mi{Sig}_k(t^{''})$ ~~with ~~$t ~~\sqsubset ~~t^{''}$; ~~or
	\item $t' ~~= ~~t_1~\oplus~\ldots~\oplus~t_n$ ~~with ~~$t  ~~\sqsubset ~~t^{''}$ ~~where ~~$t^{''} ~~\in ~~\{ t_1, ~\ldots, ~t_n \}$.

\end{itemize}

\end{definition}

We will call encrypted subterms, hashes and signatures as~~\emph{Compound Terms}; ~~we will denote them as~~$\mi{CT}(T)$~~for a set of terms,~~$T$. 

We will denote the type of a variable or atom ~~$t$ ~~as ~~$\mi{type}(t)$. We overload this to give the type of other terms. For instance, ~~$\mi{type}([t_1,\ldots,t_n])$ ~~= ~~$[\mi{type}(t_1),\ldots,\mi{type}(t_n)]$ ~~and ~~$\mi{type}([t]_k)$ ~~= ~~$[\mi{type}(t)]_{\mi{type}(k)}$. 

We will call a substitution of a term ~~$t$ ~~to a variable ~~$V$ ~~a ~~ ``well-typed"~~substitution, if ~~$\mi{type}(t) ~~= ~~\mi{type}(V)$. We will call a set of substitutions ~~$\sigma$ ~~well-typed and write ~~{\sf well-typed}$(\sigma)$ ~~if all its members are well-typed; otherwise, we call ~~$\sigma$ ~~ill-typed.

We will assume that all operators in the term algebra except the ~~\texttt{XOR} ~~operator are free of equations of the form ~~$t ~~= ~~t'$ ~~where ~~$t$ ~~and ~~$t'$ ~~are two different terms. Thus, every equation between two terms that were not constructed with the ~~\texttt{XOR}  ~~operator is of the form ~~$t ~~= ~~t$. ~~We will denote this theory, ~~$E_{\msf{STD}}$.

On the other hand, we will assume that terms created with the ~~\texttt{XOR} ~~operator to contain the following equational theory denoted ~~$E_{\msf{ACUN}}$ ~~corresponding to it's ~~{\sf ACUN} ~~algebraic properties: $t_1~~\oplus~~(t_2~~\oplus~~t_3~)$ ~~=~~$(~t_1 ~~\oplus~~t_2~)~~\oplus~~t_3$ ~~({\sf {\bf A}ssociativity});
 $t_1 ~~\oplus~~t_2$ ~~= ~~$t_2~~\oplus~~t_1$  ~~({\sf {\bf C}ommutativity}); 
  $t_1~~\oplus~~0$~~= ~~$t_1$ ~~({\sf existence of {\bf U}nity}); 
 $t_1~~\oplus~~t_1~~= ~~0$ ~~({\sf {\bf N}ilpotence}).

We will denote the unification algorithms for terms constructed purely with the standard operators and purely with the \texttt{XOR} operator as ~~$A_{\msf{STD}}$ ~~and ~~$A_{\msf{ACUN}}$ ~~respectively. 

Terms constructed using both the standard operators and the \texttt{XOR} operator can be unified using ~~$A_{\msf{STD}}$, ~~$A_{\msf{ACUN}}$ ~~and the combination algorithm of Baader \& Schulz \cite{BS96} resulting in a finite number of most general unifiers.


\subsection{Strands and Semi-bundles}\label{ss.strands-semib}

The protocol model is based on the strand space framework of \cite{THG98}.

\begin{definition} ~~{\em \bf [Node, Strand, Protocol]}\label{d.node-strand}

A ~~\emph{\bf node} ~~is a tuple ~~$\langle \mi{Sign},~ \mi{Term} \rangle$ ~~denoted ~~$+m$ ~~when it sends a term   ~~$m$, ~~or ~~$-m$ ~~when it receives ~~$m$. ~~The ~~{\em\bf sign} ~~of a node ~~$n$ ~~is denoted  ~~$\mathrm{sign}(n)$ ~~that can be ~~`$+$' ~~or ~~`$-$' ~~and its ~~{\em\bf term} ~~as ~~$\mathrm{term}(n)$ ~~derived from the term algebra.
A ~~\emph{\bf strand} ~~is a sequence of nodes denoted $\langle n_1, \ldots, n_k \rangle$ if it has $k$ nodes. Nodes in a strand are related by the edge ~~$\Rightarrow$ ~~defined such that if ~~$n_i$ ~~and ~~$n_{i+1}$ ~~belong to the same strand, then we write ~~$n_i \Rightarrow n_{i+1}$. ~~A  ~~\emph{\bf parametric strand} ~~is a strand with all parametric terms on its nodes. A ~~{\em\bf protocol} ~~is a set of parametric strands.
\end{definition}

Protocol roles (or parametric strands) can be partially instantiated to produce  ~~{\em semi-strands} ~~containing semi-terms on nodes obtained instantiating their parametric terms, depending on the knowledge of agents concerning the variables being instantiated: A variable is instantiated to an atom if the agent to which the strand corresponds to, either creates the atom according to the protocol or knows the value ({\em e.g.} being public such as an agent name). Variables may also be replaced with new variable substitutions in order for different semi-strands of the same parametric strand to be distinguishable. This is done if more than one instance of a role is visualized in an execution scenario.

We will denote the substitution to a parametric strand  ~~`$p$' ~~by an honest agent leading to a semi-strand ~~`$s$' ~~as ~~$\sigma^h_s p$.

For instance, role ~~`$A$' ~~in the ~~$\msf{NSL}_{\oplus}$  ~~protocol is the parametric strand 

\begin{center}
~~~~~~~~~~$\mbf{role_A} ~~=  ~~\langle ~~+[1,~N_A,~ A]^{\to}_{\mi{pk}(B)},  ~~-[2,~[\msf{nonce},~N_A]\oplus [\msf{agent},~A],~N_B]^{\to}_{\mi{pk}(A)},$ \\ $~~~~~~~~~~+[3,~N_B]^{\to}_{\mi{pk}(B)} ~~\rangle$
\end{center}

~~and an agent ~~`$a$' ~~that plays the role could be the semi-strand 

\begin{center}
~~~~~~~~~~$\sigma^h_{s}~\mbf{role_A} ~~= ~~\langle ~~+[1,~n_a,~a]^{\to}_{\mi{pk}(B)},  ~~-[2,~[\msf{nonce},~n_a] \oplus [\msf{agent},~a],~N_{B}]^{\to}_{\mi{pk}(a)},$ \\ $~~~~~~~~~~+[3,~N_{B}]^{\to}_{\mi{pk}(B)} ~~\rangle$
\end{center}
~~where ~~$\sigma^h_s ~~= ~~\{ a/A,~n_a/N_A \}$.

A set of semi-strands is a ~~\emph{semi-bundle}. ~~We will denote the set of all substitutions to a protocol by honest agents leading to a semi-bundle ~~$S$ ~~as $~~\sigma^H_S$. 

We will assume that honest agent substitutions leading to semi-strands are always well-typed:

\begin{assumption}\label{a.well-typed}
	Let ~~$P$ ~~be a protocol and ~~$S$ ~~be a semi-bundle such that ~~$S ~~= ~~\sigma^H_S P$. ~~Then,	~~$(\forall \sigma ~~\in ~~\sigma^H_S)(\text{\sf well-typed}(\sigma))$.	
\end{assumption}

We will use the relation ~~`{\bf precedes}' ~~($\preceq$) ~~on stand-alone strands in semi-bundles: Let ~~$s$ ~~be a strand in a semi-bundle ~~$S$. ~~Then, \\
~~$(\forall n_i, ~n_j ~~\in ~~s)(i ~~\leq ~~j ~~\Rightarrow ~~n_i ~~\preceq ~~n_j)$. 

We will abuse the notation of ~~$CT()$ ~~on strands, protocols and semi-bundles as well. We will write ~~$t ~~\in ~~S$ ~~even if ~~$t$ ~~is a term on some node of some strand of a semi-bundle ~~$S$.


\subsection{Constraints and Satisfiability}\label{ss.constraints}

We use the constraint solving model of Millen-Shmatikov \cite{MS01} that was later modified by Chevalier  \cite{C-unif2004} to model the penetrator\footnote{Heather et al. \cite{HLS03} used classical penetrator strands of \cite{THG98}, but the basic penetrator capabilities are  equal in both models.}. 

The main constraint satisfaction procedure, denoted ~~$\mbf{P_{\oplus}}$ ~~first forms a constraint sequence from an interleaving of nodes belonging to strands in a semi-bundle:

\begin{definition}~~{\em \bf [Constraint sequence]}\label{d.conseq}

A ~~{\em \bf constraint sequence} ~~$C ~~= ~~\langle  ~~\text{term}(n_1)~:~T_1,~~\ldots,~~\text{term}(n_k)~:~T_k\rangle$ ~~is from a semi-bundle ~~$S$ ~~with ~~$k$  ~~`$-$' ~~nodes if ~~$(\forall n)(\forall n')((\text{term}(n')~:~T ~~\in ~~C) \wedge (\text{term}(n) ~~\in  ~~T) \Rightarrow (n ~~\preceq ~~n'))$. ~~Further, if ~~$i < j$ ~~and ~~$n_i$, ~~$n_j$ belong to the same strand, then ~~$n_i  ~~\preceq ~~n_j$ ~~and ~~$(\forall i~=~1~\text{to}~k)(T_i ~~\subseteq  ~~T_{i+1})$.

\end{definition}

A symbolic reduction rule applied to a constraint ~~$m~:~T$  ~~is said to ``reduce" it to another constraint ~~$m~:~T'$  ~~or ~~$m'~:~T$. ~~$\mbf{P_{\oplus}}$ ~~applies a set of such rules ~~$R_{\oplus}$ ~~(Table \ref{t.rules}) in any order to the first constraint in a sequence that does not have a variable as it's target, called the ~~``{\bf active constraint}". It is worth mentioning that ~~$\mbf{P_{\oplus}}$ ~~eliminates any stand-alone or free variables in the term set of a constraint before applying any rule.

\begin{table*}
\begin{center}
\begin{tabular}{|c|c|c||c|c|c|}
	\hline 

  {\sf concat}   &   $[t_1,\ldots,t_n]~:~T$   &   $t_1~:~T$,\ldots,$t_n~:~T$   &   {\sf split}   &   $t~:~T \cup [t_1,\ldots,t_n]$	&	 $t~:~T \cup t_1 \cup \ldots \cup t_n$ \\  \hline
 
  {\sf penc}   &   $[m]^{\to}_k~:~T$   &   $k~:~T,~m~:~T$   &   {\sf pdec}   &   $m~:~[t]^{\to}_{\mi{pk}(\epsilon)}~\cup~T$	  &	 $m~:~t~\cup~T$ \\  \hline

  {\sf senc}   &   $[m]^{\leftrightarrow}_k~:~T$   &   $k~:~T,~m~:~T$   &   {\sf sdec}   &   $m~:~[t]^{\leftrightarrow}_k~\cup~T$	&	 $k~:~T,~m~:~T~\cup~\{t,k\}$ \\  \hline

  $\msf{XOR_R}$   &   $m : T \cup \{t_1, \ldots , t_n\}$   &   $m : T \cup t_1 \oplus \ldots \oplus t_n$   &   $\msf{XOR_L}$   &   $t_1 \oplus \ldots \oplus t_n : T$	  &	 $t_1 : T, t_2 \oplus \ldots \oplus t_n : T$ \\  \hline

  {\sf Sig}  & $Sig_k(f(t))~:~T$  &  $t~:~T$ &  
  {\sf Hash} & $h(t)~:~T$					&	 $t~:~T$ \\  \hline

\end{tabular}
\end{center}

\caption{Set of reduction rules, ~~$R_{\oplus} ~~= ~~R ~~\cup ~~\{ ~~\msf{XOR_L}, ~~\msf{XOR_R} ~~\}$~~}\label{t.rules}

\end{table*}

The rules in Table~\ref{t.rules} do not affect the attacker substitution. There are two other rules that involve unification, and generate a new substitution that is to be applied to the whole sequence before applying the next rule. It is worth giving a more detailed account of those rules including the transformation to the constraints before the active constraint ($C_<$) and the ones after ($C_>$):

 \[ \frac{C_<,~m : T \cup t,~C_>;~\sigma}{\tau C_<,~\tau C_>;~\tau~\cup~\sigma }~~\text{where}~\tau = \text{mgu}(m,t)~~~(\msf{un})  \]
 
 \[ \frac{ C_<,~m~:~[t]^{\to}_k~\cup~T,~C_>;~\sigma  }{ \tau~C_<,~\tau~m~:~\tau~[t]^{\to}_k~\cup~\tau~T,~\tau~C_>;~\tau~\cup~\sigma  }, 
        \text{where}~\tau = \text{mgu}(k,\mi{pk}(\epsilon)), k \neq \mi{pk}(\epsilon)~~~(\msf{ksub}) \]

A sequence of applications of reduction rules on a constraint sequence can transform it into a ~~``simple" ~~constraint sequence:

\begin{definition}~~{\em \bf [Simple constraint sequence]}\label{d.simplecons}

A constraint ~~$m~:~T$ ~~is a ~~{\em \bf simple constraint}  ~~if ~~$m$ ~~is a variable. A constraint sequence ~~$C$ ~~is a ~~{\em \bf simple constraint sequence} ~~if every constraint in ~~$C$ ~~is a simple constraint.

\end{definition}

The possibility of forming bundles from a given semi-bundle can be determined by testing if constraint sequences from it are satisfiable. Satisfiability is usually defined in terms of attacker operations on ground terms; however, Chevalier \cite{C-unif2004} proved that ~~$\mbf{P_{\oplus}}$ ~~is terminating, sound and complete with respect to the attacker capabilities. Hence, we define satisfiability directly in terms of the decision procedure:

\begin{definition}~~{\em\bf [Satisfiability]}\label{d.satisfiable}

A constraint is ~~{\em\bf satisfiable} ~~if a sequence of reduction rule application from ~~$R$ ~~result in a simple constraint. A constraint sequence ~~$C$ ~~is ~~{\em\bf  satisfiable} ~~if every constraint in the sequence is satisfiable. Further, the initially empty substitution  ~~$\sigma$ ~~is said to ~~{\em\bf satisfy} ~~$C$, ~~denoted  ~~$\sigma ~~\vdash ~~C$. 

\end{definition}

It is useful to characterize ``normal" constraint sequences which are those that do not contain pairs on the left and right sides of any constraint:

\begin{definition}~~{\em \bf [Normal Constraint Sequence]}

A constraint sequence ~~$C$ ~~is ~~{\em \bf normal} ~~iff for every constraint ~~$m~~:~~T ~~\in ~~C$, ~~$m$ ~~is not a pair and for every ~~$t ~~\in ~~T$, ~~$t$ ~~is not a pair.

\end{definition}

It has been proven in \cite{C-unif2004} that any constraint sequence can be ``normalized" such that if a substitution satisfies the original sequence, it can also satisfy the normalized sequence.

Violations of trace properties such as secrecy and authentication can be embedded in a semi-bundle so that a satisfiable constraint sequence from the semi-bundle points to an attack. Using this concept, we define a type-flaw attack: 

\begin{definition}~~{\em\bf [Type-flaw attacks]}\label{d.type-flaw}

A ~~{\em\bf type-flaw attack} ~~exists on a semi-bundle ~~$S$ ~~if a constraint sequence ~~$C$ ~~from ~~$S$ ~~is satisfiable with an ill-typed substitution, but not with a well-typed substitution. i.e. ~~$(\exists \sigma)(\sigma ~~\vdash ~~C) ~~\wedge ~~(\nexists \sigma^{'})((\sigma' ~~\vdash ~~C) \wedge (\text{\sf well-typed}(\sigma^{'})))$. 

\end{definition}

\subsection{Main requirement --- Non-Unifiability of Terms}\label{ss.NUT}

We will now state our main requirement on protocol messages  which states that textually distinct compound terms should be non-unifiable and that all XORed terms must be type-tagged:

\begin{definition}~~{\em\sf [NUT]}\label{d.NUT}

Let $P$ be a protocol. Then $P$ is ~~{\em $\msf{NUT}$-Satisfying} ~~iff 

\begin{itemize}
		\item $(\forall t_1 \in CT(P))(\forall t_2 \in 										CT(P))(t_1 \neq t_2 \Rightarrow t_1 \not\approx 						t_2).$;
		\item $(\forall t)(\forall t')((t \in P) \wedge (t = t_1 \oplus \ldots \oplus t_n) \wedge (t' \in \{ t_1, \ldots , t_n \}) \Rightarrow (\exists t^{''})(t' = [\mi{type}(t^{''}),t^{''}]))$.
		
		\end{itemize}

\end{definition}

It can be easily seen that {\sf NUT} for terms constructed with standard operators is achieved by placing component numbers as the beginning element of concatenations inside all distinct compound terms in a protocol. E.g.  $[1, ~N_A, ~A]^{\to}_{\mi{pk}(B)}$, $[2,~N_A, ~[3, ~N_B, ~A]^{\leftrightarrow}_{\mi{sh}(A,B)}]^{\leftrightarrow}_{\mi{sh}(B,S)}$, etc. Further, for terms that are XORed together, type tags must be included. For instance, ~~$N_A \oplus B \oplus [1, ~N_A, ~A]_K$ ~~should be transformed into ~~$[\msf{nonce},~N_A] \oplus [\msf{agent},~B] \oplus [[\msf{nonce},~\msf{agent}]^{\to}_{\msf{key}},~[1,~N_A,~A]^{\to}_K]$.

The tagged $\msf{NSL}_{\oplus}$ protocol in Section~\ref{s.type-flaw-ex} clearly conforms to these stipulations and hence is a {\sf NUT}-Satisfying protocol.


\subsection{Main result}\label{ss.main-result}

We will now prove that {\sf NUT}-Satisfying protocols are not vulnerable to type-flaw attacks.

The main idea is to show that every unification when applying $\mbf{P_{\oplus}}$ to a constraint sequence from a \textsf{NUT}-Satisfying protocol results in a well-typed unifier.

The intuition behind showing that unifiers are necessarily well-typed is as follows: informally, the problem of unification of two terms under the combined theory of $(E_{\msf{STD}} \cup E_{\msf{ACUN}})$ must first result in subproblems that are purely in $E_{\msf{STD}}$ or purely in $E_{\msf{ACUN}}$ according to Baader-Schulz algorithm. 

Now $E_{\msf{ACUN}}$ problems will have a unifier only if the \texttt{XOR} terms contain variables. However, according to our extended requirement of {\sf NUT} above, no protocol term has an \texttt{XOR} term with an untagged variable. Further, the \texttt{XOR} terms produced by $\mbf{P_{\oplus}}$ in the term set of a constraint cannot contain variables either since like in {\bf P}, the rule ({\em elim}) eliminates any stand-alone variables in a term set before applying any other rule. Thus, algorithm $A_{\msf{ACUN}}$ returns an empty unifier. Unification of $E_{\msf{ACUN}}$ problems only happens when two standard terms that were replaced by variables belong to the same equivalence class, can be unified with $A_{\msf{STD}}$ and could thus be canceled. 

In summary, the unifier for a problem in  $(E_{\msf{STD}} \cup E_{\msf{ACUN}})$ under the extended requirement on {\sf NUT} is only from applying $A_{\msf{STD}}$. We show that these problems always produce well-typed unifiers.

For instance, consider the unification problem

\[  [1,~n_a]_{\mi{pk}(B)} ~~\stackrel{?}{\approx_E} ~~[1,~N_B]_{\mi{pk}(a)}   ~~\oplus   ~~[2,~A] ~~\oplus  ~~[2,~b]  \]

Following Baader \& Schulz method, we first purify this to sub-problems:

\[   W ~~\stackrel{?}{\approx_{E_{\msf{STD}}}} ~~[1,~n_a]_{\mi{pk}(B)}, ~~X  ~~\stackrel{?}{{\approx_{E_{\msf{STD}}}}}  ~~[1,~N_B]_{\mi{pk}(a)}, ~~Y  ~~\stackrel{?}{\approx_{E_{\msf{STD}}}} ~~[2,~A], ~~Z  ~~\stackrel{?}{\approx_{E_{\msf{STD}}}}  ~~[2,~b], \]

\[\text{and} ~~W ~~\stackrel{?}{\approx_{E_{\msf{ACUN}}}} ~~X ~~\oplus ~~Y \oplus ~~Z.  \]

Now, the new variables ~~$W$, ~~$X$, ~~$Y$, ~~and ~~$Z$ are treated as constants during ~~$A_{\msf{ACUN}}$. ~~In that case, the problem ~~$W ~~= ~~X ~~\oplus ~~Y ~~\oplus ~~Z$ ~~is not unifiable. However, there is a step we missed: we need to form equivalence classes from the variables ~~$W$, ~~$X$, ~~$Y$, ~~and ~~$Z$ ~~such that variables from one class can be replaced with just one representative element. In this case, if we partition the variables into ~~$\{ ~\{~W~\}, ~\{~X~\}, ~\{~Y,~Z~\} ~\}$, ~~then we can change the problem $W ~~\stackrel{?}{\approx_{E_{\msf{ACUN}}}} ~~X ~~\oplus ~~Y \oplus ~~Z$ ~~into ~~$W ~~\stackrel{?}{\approx_{E_{\msf{ACUN}}}} ~~X ~~\oplus ~~Y \oplus ~~Y$ ~~with an additional problem of ~~$Y ~~\stackrel{?}{\approx_{E_{\msf{STD}}}} ~~Z$. ~~This is obviously equivalent to ~~$W ~~\stackrel{?}{\approx_{E_{\msf{STD}}}} ~~X$ ~~since the $Y$'s cancel out leading to another sub-problem.

Now all the sub-problems are purely in the ~~{\sf STD} ~~theory (terms on either sides do not involve the $\oplus$ operator):

\[  [1,~n_a]_{\mi{pk}(B)} ~~\stackrel{?}{\approx_{E_{\msf{STD}}}} ~~[1,~N_B]_{\mi{pk}(a)}, ~~[2,~A] ~~\stackrel{?}{\approx_{E_{\msf{STD}}}} ~~[2,~b].  \]

It can be easily seen that $A_{\msf{STD}}$ outputs a well-typed unifier ~~($\{ n_a/N_B, b/A \}$) ~~for these problems resulting in a well-typed unifier for a combination of $A_{\msf{STD}}$ and $A_{\msf{ACUN}}$, since $A_{\msf{ACUN}}$ outputs an empty unifier.

\begin{theorem}~~{\em \bf [\textsf{NUT} prevents type-flaw attacks]}\label{t.type-flaw}

	Let $P$ be a $\msf{NUT}$-Satisfying protocol  and $S = \sigma^H_S P$. Let $C$ be a normal constraint sequence from $S$. Then, $(\sigma \vdash C) \Rightarrow (\exists \sigma')((\sigma' \vdash C) \wedge (\text{\sf well-typed}(\sigma')))$.

\end{theorem}

\begin{proof}

If $\sigma$ satisfies $C$, then from Def.~\ref{d.satisfiable}, rules from $R_{\oplus}$ have been used to reduce it to a simple constraint sequence. The only rules that can change $\sigma$ are ($\mi{un}$) and ($\mi{ksub}$). ($\mi{ksub}$) makes a well-typed substitution since it unifies a term with the attacker's public-key which is of the same type.

We prove below that if $m~:~T \cup t \in C$, and $m \approx t$ then for each $\text{mgu}(m,t) = \tau$,  $\text{\sf well-typed}(\tau)$. Since initially $\sigma$ is empty, using induction on each constraint of the sequence, we can then conclude that $\sigma$ is well-typed.

Following the combination algorithm of \cite{BS96} described in \cite{Tuengerthal-TR-2006}, let the initial problem of $\Gamma = \{ m \stackrel{?}{\approx_E} t \}$ be reduced to $(\Gamma',<,p)$ where 

\begin{itemize}

	\item $\Gamma'$ is a set of unification problems $\{ m_1 					\stackrel{?}{\approx} t_1, \ldots, m_n 											\stackrel{?}{\approx} t_n \}$;
	\item Let $\Gamma'$ be pure with every $m 												\approx_E t \in \Gamma'$ have $m$, $t$  										formed purely from operator $\oplus$ on $0$, 								constants and variables or from the standard 								theory in the term algebra defined in 											Def.~\ref{d.term-algebra};
	\item $<$ is a linear ordering on variables such that if 					$X < Y$ then $Y$ does not occur as a subterm of the 				instantiation of $X$;
	\item $p$ is a partition $\{ V_1, V_2 \}$ on the set of 					all variables $V$ such that $V_2$ are treated as 						constants when $A_{\msf{STD}}$ is applied and $V_1$ 				are constants when $A_{\msf{ACUN}}$ is applied;
	\item Let another partition $p'$ of variables identifies 					equivalence classes of $V$ where every class in a 					partition is replaced with a representative and 						where members of the class are unifiable; 	

\end{itemize}

Let the combined unifier of $\sigma_{\msf{STD}}$ and $\sigma_{\msf{ACUN}}$ denoted $\sigma_{STD} \odot \sigma_{ACUN} = \sigma$ which is obtained by applying \cite[Def.~9]{Tuengerthal-TR-2006}; i.e., by induction on $<$. Our aim is to prove that every $\sigma$ obtained for different combinations of $<, p, p'$ is well-typed. 

Let us examine the possible forms of problem elements in $\Gamma'$:

\begin{description}

	\item[$\msf{ACUN}$ theory:] $m 
					\stackrel{?}{\approx_{\msf{ACUN}}} t$ exist where 					$m = a_1 \oplus a_2 \oplus \ldots \oplus a_i$ and 					$b = b_1 \oplus b_2 \oplus \ldots \oplus b_j$ 							where each of $a \in \{ a_1, \ldots, a_i \}$ and 						$b \in \{ b_1, \ldots, b_j \}$ is a constant or a 					new variable in $V$ for some positive $i$ and $j$. The reason is as follows: 							according to the 									requirement on 						protocol messages for a {\sf NUT}-Satisfying 								protocol, none of $\{ a_1, \ldots, a_i \}$ can be 					an untagged variable. Also, none of $\{ b_1, 								\ldots, b_j \}$ is a variable, since 												$\mbf{P_{\oplus}}$ applies rule ({\em elim}) 								eliminating all stand-alone variables before 								applying any other rule. Lastly, the new 										variables 					in $m$ and $t$ would be other 					problems in 									$\Gamma'$ of the 							form $X = 																									[\msf{tag},x]$ where $X$ is the new 												variable, $\msf{tag}$ is a constant and $x$ is 							any term. These new variables have to be treated 						as constants when applying $A_{\msf{ACUN}}$ (they cannot be substituted with 0's which is the only substitution that $A_{\msf{ACUN}}$ can return). 					
					With 					all constants in $m$ and $t$, 															$A_{\msf{ACUN}}$ that 						would normally 						return a set of `$0$' substitutions 												for some variables, returns an empty set of 								substitutions answering that $m$ and $t$ are 								equivalent (if they are); 
	
		\item[$\msf{STD}$ theory:] $m \approx_{\msf{STD}} 							t$ where 
				\begin{enumerate}
						\item either $m$ or $t$ is a new variable 												belonging to $V$; there is no unifier to 										existing variables here;

						\item $m, t$ are tagged terms of the form, 												$[\msf{tag},x]$ and $[\msf{tag},x']$ 												where $\msf{tag}$ is a constant. In this 										case, $m$ unifies with $t$ only if $x$ 											unifies with $x'$ and 	the proof can be 											applied recursively; 

						\item $m, t \in \mi{CT}(S)$; In this case, 												again the proof applies recursively. For 										instance, if $m = h(m')$ and 	$t = h(t')$ 									then we need to unify $m'$ and $t'$;
									Suppose $m' = [\msf{tag},x_1,\ldots,x_n]$ 									and $t' = [\msf{tag},y_1,\ldots,y_n]$. The constant $\msf{tag}$ guarantees that $m'$ and $t'$ have the same number of elements ($n$). Now we need to unify every $x_i$ with $y_i$ for $i = 1$ to $n$. Firstly, if one of $x_i$ and $y_i$ is a variable, then:
									\begin{itemize}
											\item	If $x_i$ or $y_i$ is a new 																	variable, there is no 																			substitution to existing 																		variables; 
											\item If both are existing variables, 														then they are both of the same 															type by Def.~\ref{d.NUT} and 																Assumption \ref{a.well-typed}; 															similarly if one of them is an 														atom;									
									\end{itemize} 						
			If $x_i, y_i \in \mi{CT}(S)$, then the proof proceeds recursively to each subterm in turn.		

 						\item $m$ and $t$ are two new variables in a 											subset of the variable identification 											partition $p'$. 																						However, since both are problems in 												$\Gamma'$ are such that they map to a 											tagged pair or compound term in the 												standard theory, their unifier is once 											again well-typed from above. Note that 											$m$ and $t$ cannot be existing variables 										since these variables are from {\sf 												ACUN} problems and {\sf ACUN} problems 											contain necessarily new variables as 												explained previously in the case for {\sf 									ACUN} theory.

				\end{enumerate}
	
\end{description}

\end{proof}

\section{More algebraic properties}\label{s.others}

We now consider some more algebraic properties of message operators. The first set breaks the free algebra assumption for protocol messages like \texttt{XOR}. The second set breaks the perfect encryption assumption.

\subsection{Algebraic properties with equational theories}

\subsubsection*{Monoidal theories.}\label{sss.monoidal} 
Following the definition of monoidal theories from \cite{CortierD07}, we can determine that 

\begin{itemize}

\item the theory {\sf ACU} over $\{ +, 0 \}$ where {\sf A} stands for associativity, {\sf C} for commutativity and {\sf U} for the existence of Unity is a monoidal theory;

\item the theories {\sf ACUIdem} and {\sf ACUN} where {\sf Idem} stands for Idempotence and {\sf N} for Nilpotence are also monoidal theories over $\{+,0\}$ and $\{\oplus,0\}$ respectively; 

\item the theory of Abelian Groups ({\sf AG} or {\sf ACUInv}) over $\{ +, -, 0 \}$ where {\sf Inv} stands for Inverse is also monoidal where $-$ is a unary operator.

\end{itemize}

If we replace or overload the $\oplus$ operator in Section \ref{s.type-flaw-proof} with {\sf Idem}  or {\sf Inv}, we can make a similar reasoning as made for {\sf ACUN} properties in Theorem \ref{t.type-flaw}: 

When the combination algorithm of Baader \& Schulz is applied for $E_{\msf{STD}} \cup E_{\msf{T}}$ where $T$ is a theory with any, some or all of {\sf A, C, U, N, Idem, Inv}, the algorithm for $\msf{T}$, say $A_{\msf{T}}$ will return an empty substitution when the operator with theory is so used in the protocol such that every term is type-tagged. Consequently, the unifier for the combined unification problem will only have substitutions from $A_{\msf{STD}}$ which will be well-typed as explained in Theorem \ref{t.type-flaw}.

However, we must note that the procedure $\mbf{P_{\oplus}}$ in \cite{C-unif2004} that we followed only considered {\sf ACUN} properties. We conjecture that if a suitable constraint solving algorithm is developed for other monoidal theories as well, then the above concept of necessarily well-typed unifiers could be used to extend Theorem \ref{t.type-flaw} under those theories.

\subsubsection{Associativity of Pairing.}
This property allows the equation $[a,[b,c]] = [[a,b],c]$. 
Denote this as the theory {\sf Assoc}.

Component numbering cannot prevents ill-typed unifiers. A simple example can prove this: $[1,A,[b,c],d]$ can be unified using $[1,[a,B],C,d]$, with $\sigma =  \{ [a,B]/A, [b,c]/C, d/D \}$. Obviously, $\sigma$ is ill-typed.

However, type-tagging prevent ill-typed unifiers. If we consider the same example, \\
$[[\msf{agent},A],[\msf{pair},[[\msf{nonce},b],[\msf{agent},c]]],[\msf{key},d]]$ cannot be unified with \\ $[[\msf{pair},[[\msf{agent},a],[\msf{nonce},B]]],[\msf{agent},C],[\msf{key},D]]$ even under associativity, due to the ``{\sf pair}'' tag for pairs. 

It would be straightforward to prove this claim formally: 

\begin{itemize}

		\item Following Baader-Schulz algorithm again, we can 						first purify the main unification problem into 							sub problems that are either purely in the {\sf 						STD} theory and through the introduction of new 						variables, to those that resemble $m \approx t$ 						where all subterms of $m$ and $t$ are variables, 						atoms or pairs for the {\sf Assoc} theory;
		\item The {\sf STD} theory returns well-typed unifiers as described in the proof of Theorem \ref{t.type-flaw}; 
		\item The unifiable problems in the {\sf Assoc} theory 						will resemble $[ [\msf{tag_1}, x_1], \ldots, 								[\msf{tag_n},x_n] ] \approx [ [\msf{tag_1}, y_1], 					\ldots, [\msf{tag_n},y_n] ]$. This returns a 								well-typed unifier if all $x_i \approx y_i$ ($i = 					1$ to $n$) return well-typed unifiers which they 						do if at least one of $x_i$ or $y_i$ are 										variables from Def. \ref{d.NUT} and Assumption 							\ref{a.well-typed}. If they are both compound 							terms, the proof proceeds recursively.

\end{itemize}

\subsubsection{Associativity and Commutativity of a general operator}

The concepts above can easily be extrapolated to associativity of a general operator, say `$.$' as well. For instance, $[1,[a.b].C]$ and $[2,A.[b.c]]$ return an ill-typed unifier, but $[[\msf{pair}, ~[\msf{agent},a].[\msf{nonce},b]].[\msf{key},C]]$ and $[[\msf{agent},A].[\msf{pair},~[\msf{nonce},b].[\msf{key},c]]]$ do not.

These concepts can be extrapolated to commutativity as well: Consider $[1,n_a.B.a]$ unified with $[1,A.b.N_A]$ that results in an ill-typed unifier $\{ n_a/A, b/B, a/N_A \}$ but type-tagging does not allow such a unification and ensures well-typed unification. Consider the same example: $[\msf{nonce}.\msf{agent}.\msf{agent},[\msf{nonce},n_a].[\msf{agent},B].[\msf{agent},a]]$ cannot be unified with  $[\msf{nonce}.\msf{agent}.\msf{agent},[\msf{agent},A].[\msf{agent},b].[\msf{nonce},N_A]]$.

It should be straightforward to extend the formal proof that we outlined for associativity of pairing to the cases of associativity and commutativity of a general operator.


\subsection{Algebraic properties with cipher weaknesses}

Some algebraic properties violate the perfect encryption assumption, without altering the freeness of the message algebra. If they produce subterms, like the following inference rule due to Coppersmith \cite{CFPR96}, the main theorem still stands tall since unification in the {\sf STD} theory will still be well-typed (recall the steps of {\sf STD} theory unification in the proof of Theorem~\ref{t.type-flaw} that handles the case of $m \approx_{\msf{STD}} t$ -- they consider $m$ and $t$ being subterms of the semi-bundle): \\

 $ \{~~[a,~x,~b]^{\rightarrow}_k,~~[c,~x,~d]^{\rightarrow}_k,~~a,~~b,~~c,~~d~~\}~~\vdash~~x, ~~\text{where}~~a \neq c ~~\vee ~~b \neq d$. \\ 

Clearly, since this inference produces a subterm ~~(`$x$'), ~~the main result stands tall in its presence and no type-flaw attacks can be possible if the protocol obeys {\sf NUT}.

Some others produce non-subterms such as the Prefix property and homomorphic encryption discussed in \cite{CDL06}. Let us examine if and how prudent tagging could be adopted to prevent type-flaw attacks under these properties:

\subsubsection*{Prefix property.} The Prefix property is obeyed by block ciphering techniques such as CBC and ECB. This property leads  the attacker to infer $[m]^{\leftrightarrow}_k$ (a non-subterm) from $[m,n]^{\leftrightarrow}_k$ thereby invalidating Theorem \ref{t.type-flaw}.

Consider the Woo and Lam $\pi_1$ protocol modified by inserting component numbers inside each encrypted component\footnote{Heather et al. \cite{HLS03} do not specify the exact position where component numbers need to be inserted, although they inserted numbers at the beginning of encryptions in their examples.}:

\begin{center}
\begin{tabular}{ll}

Msg 1. $a \to b : a$ \\
Msg 2. $b \to a : n_b$ \\
Msg 3. $a \to b : [a, b, n_b, 1]^{\leftrightarrow}_{\mi{sh}(a,s)}$ \\
Msg 4. $b \to s : [a, b, [a, b, n_b, 1]^{\leftrightarrow}_{\mi{sh}(a,s)}, 2]^{\leftrightarrow}_{\mi{sh}(b,s)}$ \\
Msg 5. $s \to b : [a, b, n_b, 3]^{\leftrightarrow}_{\mi{sh}(b,s)}$

\end{tabular}
\end{center}

$sh(x,y)$ represents a shared-key between agents $x$ and $y$. 
We presented a type-flaw attack on this protocol in \cite{MA03} even when it uses component numbering if the Prefix property is exploited, and if pairing is associative:

\begin{center}
\begin{tabular}{ll}

Msg 1. $a \to b : a$ \\
Msg 2. $b \to a : n_b$ \\
Msg 3. $I(a) \to b : [n_b, 3]$ /* In place of $[a, b, n_b, 1]_{\mi{sh}(a,s)}$ */ \\
Msg 4. $b \to I(s) : [a, b, [n_b, 3], 2]^{\leftrightarrow}_{\mi{sh}(b,s)}$ \\
Msg 5. $I(s) \to b : [a, b, n_b, 3]^{\leftrightarrow}_{\mi{sh}(b,s)}$ /* using Prefix property on Msg 4. */ \\

\end{tabular}
\end{center}

This attack works because, an attacker can infer $[a, b, n_b, 3]^{\leftrightarrow}_{\mi{sh}(b,s)}$ from Msg~4 ($[a, b, [n_b, 3], 2]^{\leftrightarrow}_{\mi{sh}(b,s)}$) exploiting the Prefix property and associativity of pairing.

This attack can be easily prevented by adopting type-tagging since it eliminates associativity of pairing as explained previously. It can also be prevented by simply inserting component numbers at the beginning of encryptions, instead of at the end.

\subsubsection*{Homomorphism of Encryptions.}
With this property, it would be possible to infer the non-subterms  $[m]_k$, and $[n]_k$ from $[m,n]_k$. Obviously, this is stronger than the Prefix property.

The ``$\msf{pair}$" tag assumed to contain within parentheses cannot void this inference. For instance, a term $[[\msf{type_1},t_1],[\msf{type_2},t_2]]_k$ can still yield the non-subterms $[\msf{type_1},t_1]_k$, and $[\msf{type_2},t_2]_k$. Even with component numbering, a term such as $[1,[t_1,t_2]]_k$ can be broken down into $[1]_k$, and $[t_1,t_2]_k$.

With a range of such non-subterm encryptions to infer, it can be easily seen that neither component numbers, nor type-tags, no matter how they are placed, can prevent the attack on the Woo and Lam protocol above under this inference. 

In particular, if the plaintext block length equals the length of a nonce or agent, then the attacker can infer $[a]_{\mi{sh}(b,s)}$, $[b]_{\mi{sh}(b,s)}$, $[n_b]_{\mi{sh}(b,s)}$ easily from Msg~4 under any tagging. He can then replay Msg~5 by stitching these together.

However, this inference is only possible under an extremely weak system such as ECB, so a realistic threat in real-world situations is unlikely.
\section{Conclusion}\label{s.conclusion}

In this paper, we provided a proof that adopting type-tagging for message fields in a protocol prevents type-flaw attacks under the {\sf ACUN} properties induced by the most popular Exclusive-OR operator. We also extrapolated those results to many other interesting and commonly encountered theories.

We did not find a single property under which component numbering prevents type-flaw attacks that type-tagging cannot, although we presented several examples where the opposite could be true. However, we advocate the use of component numbering in addition to type-tagging, since they prevent the replay of different terms with the same type as well.

The most significant advantage of being able to prevent type-flaw attacks is that analysis could be restricted to well-typed runs only. This has been shown to be a decidable problem in the standard, free theory but not for monoidal theories. We are currently in this pursuit\footnote{We previously made an attempt at this under the belief that tagging might not prevent type-flaw attacks \cite{CM07}. We intend to reattempt it by taking advantage of the results in this paper.}.

\paragraph*{Acknowledgments.} We benefited greatly from the following people's help and guidance: Gavin Lowe (OUCL) provided many useful explanations and insightful observations into type-flaw attacks during 2003-2005. Cathy Meadows (NRL) gave useful guidance and suggested adopting of Baader \& Schulz algorithm when dealing with \texttt{XOR} unification. Jon Millen (MITRE) clarified numerous concepts about constraint solving and some crucial aspects of \texttt{XOR} unification. Yannick Chevalier (IRIT) explained some concepts about his extensions to Millen-Shmatikov model with \texttt{XOR}.

\bibliographystyle{splncs}
\bibliography{SecRet-09}

\begin{thebibliography}{10}

\bibitem{HLS03}
Heather, J., Lowe, G., Schneider, S.:
\newblock How to prevent type flaw attacks on security protocols.
\newblock Journal of Computer Security \textbf{11}(2) (2003)  217--244

\bibitem{THG98}
Thayer, F.J., Herzog, J.C., Guttman, J.D.:
\newblock Strand spaces: Why is a security protocol correct?
\newblock In: Proc. IEEE Symposium on Research in Security and Privacy, IEEE
  Computer Society Press (1998)  160--171

\bibitem{Cohen00}
Cohen, E.:
\newblock Taps: A first-order verifier for cryptographic protocols.
\newblock In: Computer Security Foundations Workshop (CSFW). (2000)  144--158

\bibitem{HS00}
Heather, J., Schneider, S.:
\newblock Towards automatic verification of security protocols on an unbounded
  network.
\newblock In: Proc. 13th Computer Security Foundations Workshop, IEEE Computer
  Society Press (2000)  132--143

\bibitem{RS-FST03}
Ramanujam, R., Suresh, S.P.:
\newblock Tagging makes secrecy decidable for unbounded nonces as well.
\newblock In: 23rd FST\&TCS, Lecture Notes in Computer Science. Volume 2914.
  (2003)  323--374

\bibitem{KustersT08}
K{\"u}sters, R., Truderung, T.:
\newblock Reducing protocol analysis with xor to the xor-free case in the horn
  theory based approach.
\newblock In: ACM Conference on Computer and Communications Security. (2008)
  129--138

\bibitem{EscobarMM07}
Escobar, S., Meadows, C., Meseguer, J.:
\newblock Equational cryptographic reasoning in the maude-nrl protocol
  analyzer.
\newblock Electr. Notes Theor. Comput. Sci. \textbf{171}(4) (2007)  23--36

\bibitem{CDL06}
Cortier, V., Delaune, S., Lafourcade, P.:
\newblock A survey of algebraic properties used in cryptographic protocols.
\newblock Journal of Computer Security \textbf{14}(1) (2006)  1--43

\bibitem{CKRT03}
Chevalier, Y., K\"{u}sters, R., Rusinowitch, M., Turuani, M.:
\newblock {An NP decision procedure for protocol insecurity with XOR}.
\newblock In: Proc. 18$^{th}$ Annual IEEE Symposium on Logic in Computer
  Science (LICS'03), IEEE Computer Society Press (2003)  261--270

\bibitem{MH08}
Malladi, S., Hura, G.S.:
\newblock What is the best way to prove a cryptographic protocol correct?
  (position paper).
\newblock In: Workshop on Security in Systems and Networks (SSN 2008), IEEE
  International Symposium on Parallel and Distributed Processing (IPDPS 2008).
  (2008)  1--7

\bibitem{Mall04}
Malladi, S.:
\newblock Phd dissertation - formal analysis and verification of password
  protocols.
\newblock ACM Portal, University of Idaho (2004)

\bibitem{BS96}
Baader, F., Schulz, K.U.:
\newblock Unification in the union of disjoint equational theories: Combining
  decision procedures.
\newblock J. of Symbolic Computation \textbf{21} (1996)  211--243

\bibitem{MS01}
Millen, J., Shmatikov, V.:
\newblock {Constraint solving for bounded-process cryptographic protocol
  analysis}.
\newblock In: Proc. ACM Conference on Computer and Communication Security, ACM
  press (2001)  166--175

\bibitem{C-unif2004}
Chevalier, Y.:
\newblock A simple constraint solving procedure for protocols with
  exclusive-or.
\newblock Presented at Unif 2004 workshop (2004) available at
  http://w\-ww.lsv.e\-ns-cacha\-n.fr/un\-if/past/un\-if04/pro\-gram.ht\-ml.

\bibitem{Tuengerthal-TR-2006}
Tuengerthal, M.:
\newblock {Implementing a Unification Algorithm for Protocol Analysis with
  XOR}.
\newblock Technical Report 0609, Institut f{\"u}r Informatik, CAU Kiel, Germany
  (2006)

\bibitem{CortierD07}
Cortier, V., Delaune, S.:
\newblock Deciding knowledge in security protocols for monoidal equational
  theories.
\newblock In: LPAR. (2007)  196--210

\bibitem{CFPR96}
Coppersmith, D., Franklin, M., Patarin, J., Reiter, M.:
\newblock {Low-exponent RSA with related messages}.
\newblock Lecture notes in computer science \textbf{1070} (1996)

\bibitem{MA03}
Malladi, S., Alves-Foss, J.:
\newblock {How to prevent type-flaw guessing attacks on password protocols}.
\newblock In: Workshop on Foundations of Computer Security (FCS03), Ottawa,
  Canada (2003)

\bibitem{CM07}
Chevalier, Y., Malladi, S.:
\newblock Decidability of ``real-world" context-explicit security protocols.
\newblock Unpublished draft (2007)

\end{thebibliography}


\begin{thebibliography}{CKRT03}

\bibitem[AN94]{AN94}
M.~Abadi and R.~Needham.
\newblock {Prudent Engineering Practice for Cryptographic Protocols}.
\newblock In {\em Proc. IEEE Symposium on Research in Security and Privacy},
  pages 122--136. IEEE Computer Society Press, 1994.

\bibitem[BS96]{BS96}
F.~Baader and K.~U. Schulz.
\newblock Unification in the union of disjoint equational theories: Combining
  decision procedures.
\newblock {\em J. of Symbolic Computation}, 21:211--243, 1996.

\bibitem[CD07]{CortierD07}
V{\'e}ronique Cortier and St{\'e}phanie Delaune.
\newblock Deciding knowledge in security protocols for monoidal equational
  theories.
\newblock In {\em LPAR}, pages 196--210, 2007.

\bibitem[CDL06]{CortierDL06}
V{\'e}ronique Cortier, St{\'e}phanie Delaune, and Pascal Lafourcade.
\newblock A survey of algebraic properties used in cryptographic protocols.
\newblock {\em Journal of Computer Security}, 14(1):1--43, 2006.

\bibitem[CFPR96]{CFPR96}
D.~Coppersmith, M.~Franklin, J.~Patarin, and M.~Reiter.
\newblock {Low-exponent RSA with related messages}.
\newblock {\em Lecture notes in computer science}, 1070, 1996.

\bibitem[Che04]{C-unif2004}
Y.~Chevalier.
\newblock A simple constraint solving procedure for protocols with
  exclusive-or.
\newblock {\em Presented at Unif 2004 workshop}, 2004.
\newblock available at
  http://w\-ww.lsv.e\-ns-cacha\-n.fr/un\-if/past/un\-if04/pro\-gram.ht\-ml.

\bibitem[CKRT03]{CKRT03}
Y.~Chevalier, R.~K\"{u}sters, M.~Rusinowitch, and M.~Turuani.
\newblock {An NP decision procedure for protocol insecurity with XOR}.
\newblock In {\em Proc. 18$^{th}$ Annual IEEE Symposium on Logic in Computer
  Science (LICS'03)}, pages 261--270. IEEE Computer Society Press, 2003.

\bibitem[Coh00]{Cohen00}
E.~Cohen.
\newblock Taps: A first-order verifier for cryptographic protocols.
\newblock In {\em Computer Security Foundations Workshop (CSFW)}, pages
  144--158, 2000.

\bibitem[EMM07]{EscobarMM07}
Santiago Escobar, Catherine Meadows, and Jos{\'e} Meseguer.
\newblock Equational cryptographic reasoning in the maude-nrl protocol
  analyzer.
\newblock {\em Electr. Notes Theor. Comput. Sci.}, 171(4):23--36, 2007.

\bibitem[HLS00]{HLS00}
J.~Heather, G.~Lowe, and S.~Schneider.
\newblock How to prevent type flaw attacks on security protocols.
\newblock In {\em Proc. 13th Computer Security Foundations Workshop}, pages
  255--268. IEEE Computer Society Press, July 2000.

\bibitem[HLS03]{HLS03}
J.~Heather, G.~Lowe, and S.~Schneider.
\newblock How to prevent type flaw attacks on security protocols.
\newblock {\em Journal of Computer Security}, 11(2):217--244, 2003.

\bibitem[HS00]{HS00}
J.~Heather and S.~Schneider.
\newblock Towards automatic verification of security protocols on an unbounded
  network.
\newblock In {\em Proc. 13th Computer Security Foundations Workshop}, pages
  132--143. IEEE Computer Society Press, 2000.

\bibitem[KT08]{KustersT08}
Ralf K{\"u}sters and Tomasz Truderung.
\newblock Reducing protocol analysis with xor to the xor-free case in the horn
  theory based approach.
\newblock In {\em ACM Conference on Computer and Communications Security},
  pages 129--138, 2008.

\bibitem[MAF03]{MA03}
S.~Malladi and J.~Alves-Foss.
\newblock {How to prevent type-flaw guessing attacks on password protocols}.
\newblock In {\em Workshop on Foundations of Computer Security (FCS03)},
  Ottawa, Canada, June 2003.

\bibitem[MH08]{MH08}
S.~Malladi and G.~S. Hura.
\newblock What is the best way to prove a cryptographic protocol correct?
  (position paper).
\newblock In {\em Workshop on Security in Systems and Networks (SSN 2008), IEEE
  International Symposium on Parallel and Distributed Processing (IPDPS 2008)},
  pages 1--7, 2008.

\bibitem[MR05]{MR05}
S.~Malladi and S.~Rosenberg.
\newblock Extending constraint solving for cryptographic protocol analysis with
  non-standard attacker inference rules.
\newblock In {\em IASTED International Conference on Communication, Network and
  Information Security (CNIS 2005)}, November 2005.

\bibitem[MS01]{MS01}
J.~Millen and V.~Shmatikov.
\newblock {Constraint solving for bounded-process cryptographic protocol
  analysis}.
\newblock In {\em Proc. ACM Conference on Computer and Communication Security},
  pages 166--175. ACM press, 2001.

\bibitem[RS03]{RS-FST03}
R.~Ramanujam and S.~P. Suresh.
\newblock Tagging makes secrecy decidable for unbounded nonces as well.
\newblock In {\em 23rd FST\&TCS, Lecture Notes in Computer Science}, volume
  2914, pages 323--374, December 2003.

\bibitem[THG98]{THG98}
F.~J. Thayer, J.~C. Herzog, and J.~D. Guttman.
\newblock Strand spaces: Why is a security protocol correct?
\newblock In {\em Proc. IEEE Symposium on Research in Security and Privacy},
  pages 160--171. IEEE Computer Society Press, 1998.

\bibitem[Tue06]{Tuengerthal-TR-2006}
M.~Tuengerthal.
\newblock {Implementing a Unification Algorithm for Protocol Analysis with
  XOR}.
\newblock Technical Report 0609, Institut f{\"u}r Informatik, CAU Kiel,
  Germany, 2006.

\end{thebibliography}


\end{document}